\documentclass[reprint,amsmath,amssymb,floatfix,aps, superscriptaddress]{revtex4-2}
\usepackage[version=3]{mhchem} % Formula subscripts using \ce{}
\usepackage{graphicx}
\usepackage{enumitem}
\usepackage{xcolor}
\usepackage[normalem]{ulem}
\usepackage[fontsize=10pt]{scrextend}
\usepackage{graphicx}
\usepackage{wrapfig}
\usepackage{framed}
\usepackage{color}
\usepackage{float}
\usepackage{multirow}
\usepackage{amsmath} 
\usepackage{soul}
\makeatletter

\newcommand*{\rom}[1]{\expandafter\@slowromancap\romannumeral #1@}
\makeatother

\newcommand{\etal}{\textit{et al. }}

\begin{document}
\preprint{APS/123-QED}
%%%%%%%%%%%%%%%%%%%%%%%%%%%%%%%%%%%%%%%%%%%%%%%%%%
\title{Emergence of Order in Chemically Active Droplets: Temporal Dynamics and Collective Behavior}
%%%%%%%%%%%%%%%%%%%%%%%%%%%%%%%%%%%%%%%%%%%%%%%%%%
\author{Sobiya Ashraf}
\affiliation{Department of Chemical Engineering, Indian Institute of Technology Kanpur, Kanpur, India.}

\author{Pawan Kumar}
\affiliation{Department of Chemical Engineering, Indian Institute of Technology Kanpur, Kanpur, India.}

\author{Prateek Dwivedi}
\affiliation{Department of Chemical Engineering, Indian Institute of Technology Kanpur, Kanpur, India.}

\author{Frédéric Blanc}
\affiliation{Institut de Physique de Nice, CNRS Université Côte d’Azur, France}

\author{Dipin Pillai}
\affiliation{Department of Chemical Engineering, Indian Institute of Technology Kanpur, Kanpur, India.}

\author{Rahul Mangal}
\email{mangalr@iitk.ac.in}
\affiliation{Department of Chemical Engineering, Indian Institute of Technology Kanpur, Kanpur, India.}

%%%%%%%%%%%%%%%%%%%%%%%%%%%%%%%%%%%%%%%%%%%%%%%%%%%%%%%%%%%

\begin{abstract}

%Active droplets represent excellent synthetic mimics to study a number of intricate phenomena resulting in a population of biological microswimmers viz chemotaxis ,active turbulence and navigation in a structured environment. Here, we investigated the collective behavior of 4-Cyano-4'-pentyl-biphenyl (5CB) oil droplets under the influence of Péclet ($Pe$). At high Pe, droplets swim with pusher mode and form metastable chain-like clusters. The decrease in $Pe$ values is found to enhance the repulsive interactions among the droplets, which suppresses the clustering. Interestingly, at low Pe regime, droplets self-organise into a crystalline-like state with hexatic state.  Finally, we show the navigation of larger size droplets in the structured environment and found a significant enhancement in their swimming speed.
Collective behaviors such as swarming, chemical signaling, and clustering are fundamental to biological microorganisms, enabling hierarchical colony formation, coordinated motion, and enhanced nutrient accessibility—crucial for their survival. Over the past few decades, extensive research has been dedicated to unraveling the mechanisms underlying these diverse collective patterns through experimental model systems. Among these, active droplets have emerged as valuable synthetic analogs, effectively replicating key biological attributes and serving as ideal platforms for investigating collective phenomena. This research explores the collective behavior of 4-Cyano-4'-pentyl-biphenyl (5CB) oil droplets across varying Péclet ($Pe$) numbers. At high $Pe$, droplets exhibit a pusher mode of propulsion and form dynamic chain-like patterns. Decreasing $Pe$ enhances repulsive interactions among droplets, resulting in the inhibition of clustering. In the low $Pe$ regime, their repulsive interactions predominated by chemical field lead to the emergence of an ordered structure. Furthermore, we illustrate how active droplets efficiently navigate within a soft structured environment. These findings contribute to our comprehension of self-organized phenomena in active matter systems and provide insights for designing strategies for controlled locomotion in intricate fluidic environments.

\end{abstract}
%%%%%%%%%%%%%%%%%%%%%%%%%%%%%%%%%%%%%%%%%%%%%%%%%%%%%%%

\maketitle

\section{Introduction}
Microorganisms, due to their micron-scale dimensions, operate at low Reynolds numbers compared to macroscopic beings. As a result, even in water—a medium of relatively low viscosity, they are subject to pronounced viscous forces, requiring unique mechanisms to achieve self-propulsion. To maneuver in environments characterized by low Reynolds numbers, these organisms utilize non-time-reversible (or asymmetric) strokes \cite{purcell2014life}, breaking symmetry through mechanisms such as the coordinated, non-reciprocal motion of cellular appendages like cilia and flagella, or by deforming their body shape.

To emulate the movement of biological swimmers, researchers have developed artificial microswimmers that replicate the actions of cilia and flagella. These artificial swimmers achieve self-propulsion either through externally powered mechanisms \cite{moran2017phoretic} or by harnessing chemical energy to induce asymmetric movement in mechanically active components, a key requirement for propulsion. However, self-propulsion via mechanically moving parts often faces challenges in achieving truly autonomous motion, as it typically relies on external forces. To overcome these limitations, researchers have explored the development of artificial chemical microswimmers, which utilize chemical reactions to generate interfacial flows, driving their motion. These microswimmers operate autonomously without the need for external mechanical forcing, making them a promising advancement in fields such as micro-robotics and targeted drug delivery.

Among artificial chemical microswimmers, one of the most commonly studied are the ones based on Janus Particles (JPs) which are propelled via a mechanism known as ``phoresis'' \cite{anderson1989colloid}. In this mechanism a colloidal particle moves in response to gradients in a physical quantity. JPs exploit their intrinsic morphological asymmetry to generate a gradient of physical quantities such as chemical concentration, temperature, and ion concentration, across their surfaces. Since the gradient is generated by the particles themselves, the resulting propulsion mechanism is termed self-diffusiophoresis \cite{howse2007self, golestanian2007designing}, self-thermophoresis \cite{jiang2010active, yang2011simulations}, or self-electrophoresis \cite{paxton2004catalytic, wang2006bipolar}, depending on whether the motion is driven by local gradients in solute concentration, temperature, or ionic fields, respectively.

The other class of artificial chemical micro-swimmers consists of active droplets, which, in contrast to Janus particles, are inherently isotropic. These droplets rely on either interfacial chemical reaction \cite{thutupalli2011swarming,kumar2021fast} or micellar solubilization \cite{dwivedi2022self, peddireddy2012solubilization, michelin2023self, birrer2022we} to spontaneously break the symmetry of their interfacial tension, enabling net motion without requiring any pre-existing structural asymmetry. The resulting interfacial gradient generates a Marangoni flow, which drives the fluid at the interface and causes the droplet to move in the opposite direction, thereby conserving linear momentum and satisfying the force-free condition. Due to its simplicity, micellar solubilization is the most commonly. Briefly, this mechanism facilitates the local transfer of dispersed phase molecules from the droplet to free micelles, thereby generating filled micelles. Since filled micelles are larger in size compared to the empty micelles, this process is accompanied by a depletion of surfactants from the droplet interface. The resulting asymmetry at the droplet interface drives the fluid from the low interfacial tension region towards the high interfacial tension region, which in turn propels the droplet in the opposite direction \cite{herminghaus2014interfacial}. Once the droplet begins to move, it constantly encounters empty micelles at the front and leaves behind a trail of filled micelles at the rear, which are low in fresh surfactant molecules. This process maintains the interfacial tension gradient and thus sustains the droplet's self-propulsion. The timescale associated with the advection of empty micelles and the diffusion of filled micelles into the surrounding medium play a crucial role in determining the motion characteristics of the droplets. The relative importance of these two timescales, captured by the Péclet number ($Pe$), governs the mode of droplet motion \cite{morozov2019nonlinear}. Changes in droplet size and the addition of solutes, both molecular and macromolecular, have been shown to alter the Péclet number, leading to transitions in their swimming mode. As $Pe$ increases, the droplet's motion shifts from smooth persistent to random, and eventually to a jittery behavior \cite{suda2021straight,dwivedi2023mode, dwivedi2021solute,hokmabad2021emergence}. The change in the nature of trajectories is attributed to a transition in the swimming mode, evolving from an asymmetric dipolar puller to a pusher, and finally to a combination of a pusher with symmetric quadrupolar stationary modes. 

%From low to high Péclet number, the droplets have been shown to perform a straight to curvilinear motion \cite{suda2021straight,dwivedi2023mode}. The presence of solutes in the surrounding medium leads to significant altercation in the motion of the droplets as was evident with the emergent  jitteriness of the motion \cite{dwivedi2021solute,hokmabad2021emergence} upon adding glycerol ,smooth persistent motion  upon adding large molecular weight polymer-PEO\cite{dwivedi2023mode},and less curling motion with puller mode of swimming upon adding filled micelles \cite{kumar2023emergent}. 

Regarding their individual motion, active droplets have been shown to exhibit several intruiguing behaviors, including chemotaxis \cite{jin2017chemotaxis,hokmabad2022chemotactic}, rheotaxis in external flows \cite{dwivedi2021rheotaxis,dey2022oscillatory}, shape deformation in viscoelastic media \cite{dwivedi2023deforming}, and obstacle sensing, which can lead to orbiting or trapping around pillars \cite{jin2019fine}. Beyond these fascinating solitary behaviors, active droplets also exhibit intriguing pair interactions influenced by activity and confinement geometry, such as predator-prey-like chasing \cite{meredith2020predator, kumar2024motility} and side-wise guiding \cite{hokmabad2022spontaneously}. Recently Dwivedi {\it{et al.}} \cite{dwivedi2024p} observed that tuning the $Pe$ of self-propelled 4-Cyano-4'-pentylbiphenyl (5CB) droplets in quasi-2D confinement revealed distinct pairwise interactions. In conventional aqueous surfactant solutions, offering moderate $Pe$ droplets exhibited side-wise attraction and come in physical contact, driven by their hydrodynamic signature as being pushers.  However, when high molecular weight (8000 kDa) polyethylene oxide (PEO) was introduced into the surfactant medium, leading to a lower $Pe$, the droplets underwent a behavioral transformation. Despite adopting a puller swimming mode, they consistently exhibited long-range repulsive chemical interactions, avoiding physical contact.

In terms of their collective motion, they have been demonstrated to display highly complex collective motion$/$dynamic self-assemblies , which are influenced by various factors such as system dimensionality, the nature of confining boundaries, and the droplets' intrinsic activity \cite{thutupalli2018flow, krueger2016dimensionality, hokmabad2022spontaneously, kumar2024emergent}. For example, depending on the height of the optical cell, droplets can form a range of structures, from dynamic, arch-shaped chain-like assemblies in quasi-2D confinement to stable traveling bands at intermediate reservoir heights. At unconfined heights, the system can stabilize into large hexagonally packed clusters driven by convection \cite{thutupalli2018flow, krueger2016dimensionality}. Notably, when one of the confining boundaries is replaced with an air interface, the suppression of the stabilizing convection rolls results in the transition of large hexagonal clusters into small dynamic 2D hexagonal crystallites.

Despite these fascinating behaviors, most research on the collective dynamics of active droplets has concentrated on systems with moderate to high Péclet numbers, where hydrodynamic forces dominate droplet motion and inter-droplet interactions. In such systems, chemical effects often play a secondary role, overshadowed by the hydrodynamic signature of the droplets. Although chemotactic self-caging due to droplet-trail interactions has been reported \cite{hokmabad2022chemotactic}, the role of direct chemical interactions between droplets—expected to emerge at lower Péclet numbers—remains largely unexplored. Recently, Liu \emph{et al}. \cite{Liu2024Self-Organized} demonstrated that chemically active droplets can form diverse self-organized structures driven by attractive-repulsive chemical interactions, influenced by droplet number ratios and relative droplet diameters. Such insights into collective behaviors while being governed purely by chemical interactions remain largely restricted to non-reciprocal systems, and have been observed over relatively short time scales.

Therefore, in this report, we investigate the collective behavior of micellar solubilization-based identical active 4-cyano-4'-pentylbiphenyl (5CB) droplets in an aqueous tetradecyltrimethylammonium bromide (TTAB) solution, over an extended duration ($\sim$ 3 hours), both with and without polyethylene oxide (PEO) as a macromolecular solute, in a quasi-2D confinement. The presence of PEO allows us to modulate the underlying Péclet number ($Pe$) of the droplets. Through carefully designed experiments, we demonstrate that active droplets exhibit distinct behaviors across two Péclet regimes—with hexatic order resembling Wigner crystals emerging at low Péclet ($Pe$) numbers, while a random droplet distribution is observed at higher $Pe$. Additionally, we show that this ordered arrangement of repulsive droplets creates a chemically structured environment, continuously emitting solutes, which guides the motion of a larger active droplet. This study offers valuable insights into pattern formation in out-of-equilibrium life-like systems and applicable strategies to realize such assemblies. 
%The type of hexatic ordering has not been reported in self-propelled droplets but has been previously reported for some active systems via experiments\cite{soh2008dynamic}and via simulations\cite{gouiller2021two,yang2024shaping} 

%akin to the motion of bacteria in ordered porous media. \cite{dehkharghani2023self}

\section{Materials and Methods}
A Polydimethylsiloxane (PDMS) well with a depth of $h \sim$ 100 µm and a diameter of 5.2 {$\pm$ 0.1} mm was fabricated using a 20:1 ratio of silicone elastomer (Sylgard) and its curing agent (Sylgard). The well was thoroughly cleaned, first by washing it with 50 vol.$\%$ ethanol, followed by drying with nitrogen gas. The well was filled with aqueous solution containing 6 wt.$\%$ tetradecyl-trimethyl-ammonium bromide (TTAB, 96$\%$, Loba Chemicals) as a surfactant, with and without 1 wt.$\%$ polyethylene oxide (PEO, 8000 kDa, Sigma Aldrich) as a macromolecular additive. All of the solutions were prepared using ultrapure water (Smart2Pure 6 UV/UF). 

\begin{figure}[t]
  {\includegraphics[scale=0.53]{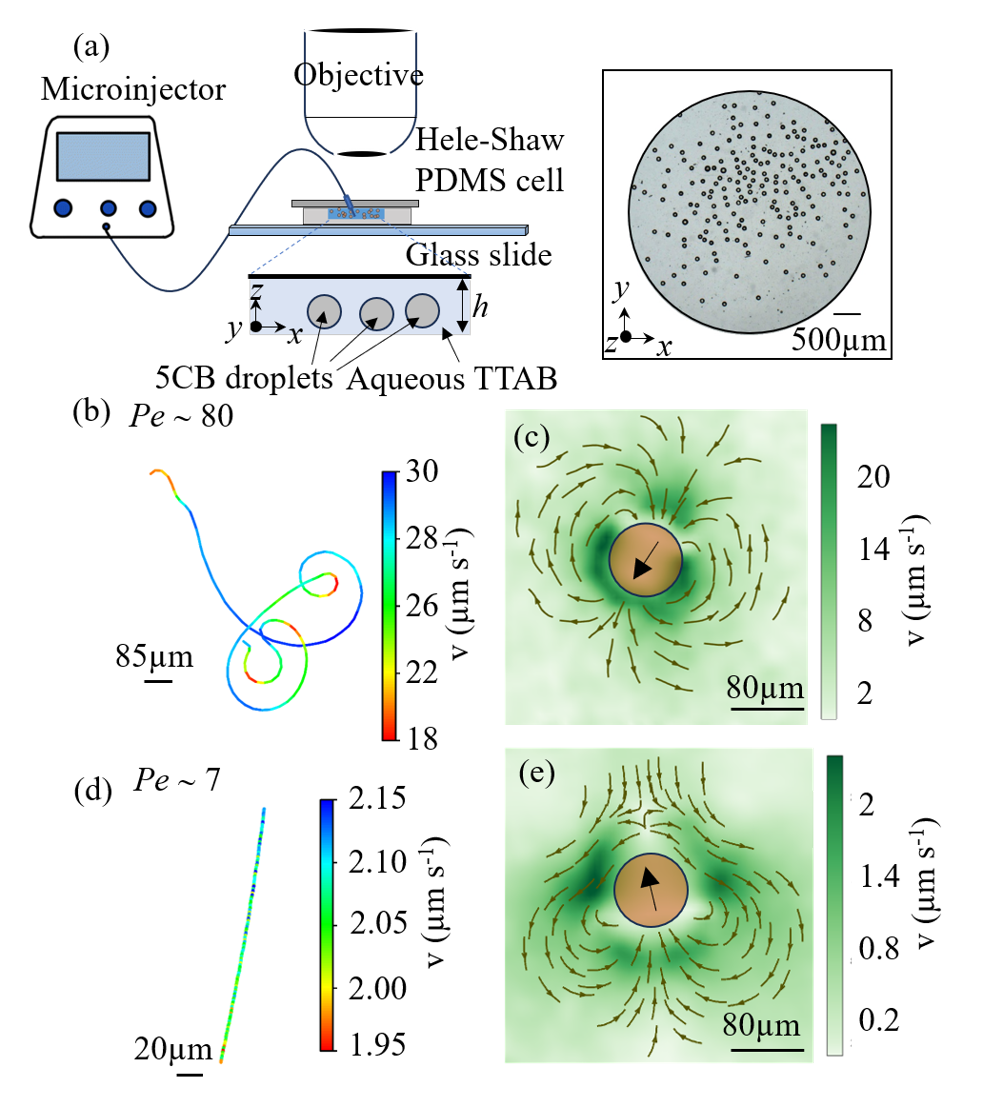}}
\linespread{1.0}
\caption{\small (a) Schematic depicting the experimental setup, and a representative optical micrograph illustrating the $x-y$ view of the cell. (b) Representative $x-y$ trajectory (duration $\Delta{\textit{t}}$ = 120 s) of the swimming 5CB droplet in a 6 wt.$\%$ TTAB aqueous solution, and corresponding (c) PIV micrograph with flow streamlines around the swimming droplet. (d) Representative $x-y$ trajectory (duration $\Delta{\textit{t}}$ = 120 s) of the 5CB droplet in 6 wt.$\%$ TTAB aqueous solution containing 1 wt.$\%$ PEO, and corresponding (e) PIV micrograph with flow streamlines around the droplet. }
\label{fig1}
\end{figure}

Using microinjector (Femtojet 4i, Eppendorf), we introduced $\sim$ 200-600 micro-droplets of 4-Cyano-4'-pentylbiphenyl (5CB, Jiangsu Hecheng Advanced Materials), each with a diameter of $\sim$ 85 µm, into the PDMS well. Subsequently, the well was sealed with a borosilicate microscope cover glass. Considering the size of the droplets to be nearly similar to the well depth, the droplets were confined to move in a 2-dimensional setting, as illustrated in Figure \ref{fig1}a. To track the droplets' center of mass position (x(t), y(t)) and their displacement with time, bright-field experiments were conducted by mounting the PDMS well on the stage of an upright microscope Olympus BX53. The motion of the droplets was recorded with an Olympus BFS-U3-70S7C-C FLIR camera at 1 frame/s in bright field mode. All experiments were carried out at fixed temperature of 25$\pm$1 $^o$C.

%%%%%%%%%%%%%%%%%%%%%%%%%%%%%%%%%%%%%%%%%%%%%%%%
%%%%%%%%%%%%%%%%%%%%%%%%%%%%%%%%%%%%%%%%%%%%%%%

To capture the flow field around the droplets, we employed particle image velocimetry (PIV). Red fluorescent polystyrene microspheres (1000 nm, Thermo Scientific) as tracers were dispersed in the aqueous TTAB solution (with and without PEO). The optical cell was positioned on an inverted microscope Olympus IX73 connected with a fluorescence illuminator Olympus U-RFL-T (Mercury Burner USH-1030L), with a laser ($\lambda$ $\sim$ 560 nm) to excite the red dye molecules. The fluorescence videos were captured with an ORX-10G-71S7C-C FLIR camera attached to the microscope and monitored via SpinView software using constant exposure throughout the experiments. For characterizing the chemical field around the droplets, 5CB droplets were doped with oil-soluble fluorescent dye (Nile Red, Sigma Aldrich) and then subjected to fluorescence microscopy to capture the distribution of the released chemical field from the active droplets.The x-y trajectories of droplets were analyzed using ImageJ software with the Trackmate \cite{ershov2022trackmate} and MultiTracker plugins. The velocity vector flow field data around the droplets were determined using PIVlab open-source package in MATLAB \cite{thielicke2021particle}, and streamlines were plotted in Techplot360. For measuring the zero shear viscosity, a viscometer (ROTAVISC lo-vi, IKA)  with a concentric cylinder setup (ELVAS-1, IKA) was used.

\section{Results and discussion}
\subsection{Isolated droplet dynamics}
We started with the exploration of self-propelled motion of an isolated droplet of 5CB introduced into an aqueous solution of 6 wt.$\%$ TTAB. For surfactant concentration significantly higher than its critical micellar concentration (CMC) $\sim$ 0.13 wt.$\%$, droplets are known to perform self-propulsion through micellar solubilization \cite{peddireddy2012solubilization, todorov2002kinetics, dwivedi2022self, herminghaus2014interfacial}. The representative $x-y$ trajectory of a 5CB droplet in a 6 wt.\% TTAB aqueous solution, depicted in Figure \ref{fig1}b and color-coded to indicate speed, illustrates the anticipated curling behavior \cite{kruger2016curling}, with the droplet swimming at an average speed $v$ of 25 $\mu$m s$^{-1}$. The significance of fluid flow field and distribution of chemical field (of filled micelles) surrounding the droplet is quantified by estimating the Péclet number ($Pe$), which is defined as the ratio of advection to diffusion of filled micelles, given by $Pe=\frac{av}{D}$, where $a$ denotes the droplet diameter and $D$ is the diffusivity of filled micelles. The determination of $D$ involved visualizing the expansion of released chemical trail perpendicular to the droplet's swimming direction at a distance of 2$a$ from the droplet's centroid. Gaussian fits of normalized intensity of filled micelles (supporting Figure S1(a,b)) reveal that the spatial variance ($\sigma{^2}$) increases linearly with time ($\Delta{t}$), as described by $\sigma{^2}=2D\Delta{t}$. From this, we obtained $D\sim 28\pm5$  $\mu$m$^2$ s$^{-1}$, consistent with previous findings by Dwivedi {\it{et al.}} \cite{dwivedi2023mode}. These measurements yield a Péclet number of $Pe \sim 80$. This high $Pe$ value, combined with the nematic phase of the 5CB droplet, is expected to contribute to its curling trajectory during motion \cite{kruger2016curling, suga2018self}. Furthermore, PIV characterization of fluid flow field (Figure \ref{fig1}c) and the associated tangential velocity along the droplet periphery when fitted using squirmer model \cite{blake1971spherical} reveal a negative swimming parameter ($\beta$ $\sim$ -0.3), characteristic of a `weak-pusher' mode of swimming (see supporting Figure S1(c)).\\

Upon introducing 1 wt.$\%$ PEO into 6 wt.$\%$ TTAB solution, the corresponding $Pe$ significantly decreases to 7, resulting in droplets exhibiting straight-line motion, as depicted in Figure \ref{fig1}d. This reduction in $Pe$ is primarily attributed to the larger size of the PEO in the surrounding phase compared to the filled micelle, producing distinct viscosity effects for the fluid advection (governed by the bulk viscosity $\eta_{bulk}$= 11 Pa-s) vs the diffusion of the chemical field (governed by local viscosity $\eta_{local}$ $\ll$ $\eta_{bulk}$) \cite{dwivedi2023mode}. As a consequence, upon addition of PEO, the droplets' propulsion speed is reduced, while the micelle diffusivity remains largely unchanged. he enhanced persistence of motion is attributed to the fluid flow field (Figure \ref{fig1}e), which is characteristic of a puller swimming mode, confirmed by the assessment of the swimming parameter $\beta \sim 0.88$ (see supporting Figure S1(c)).

%%%%%%%%%%%%%%%%%%%%%%%%%%%%%%%%%%%%%%%%%%%%%%%%%%%%%
\begin{figure*}
    \centering
    \includegraphics[width=1\linewidth]{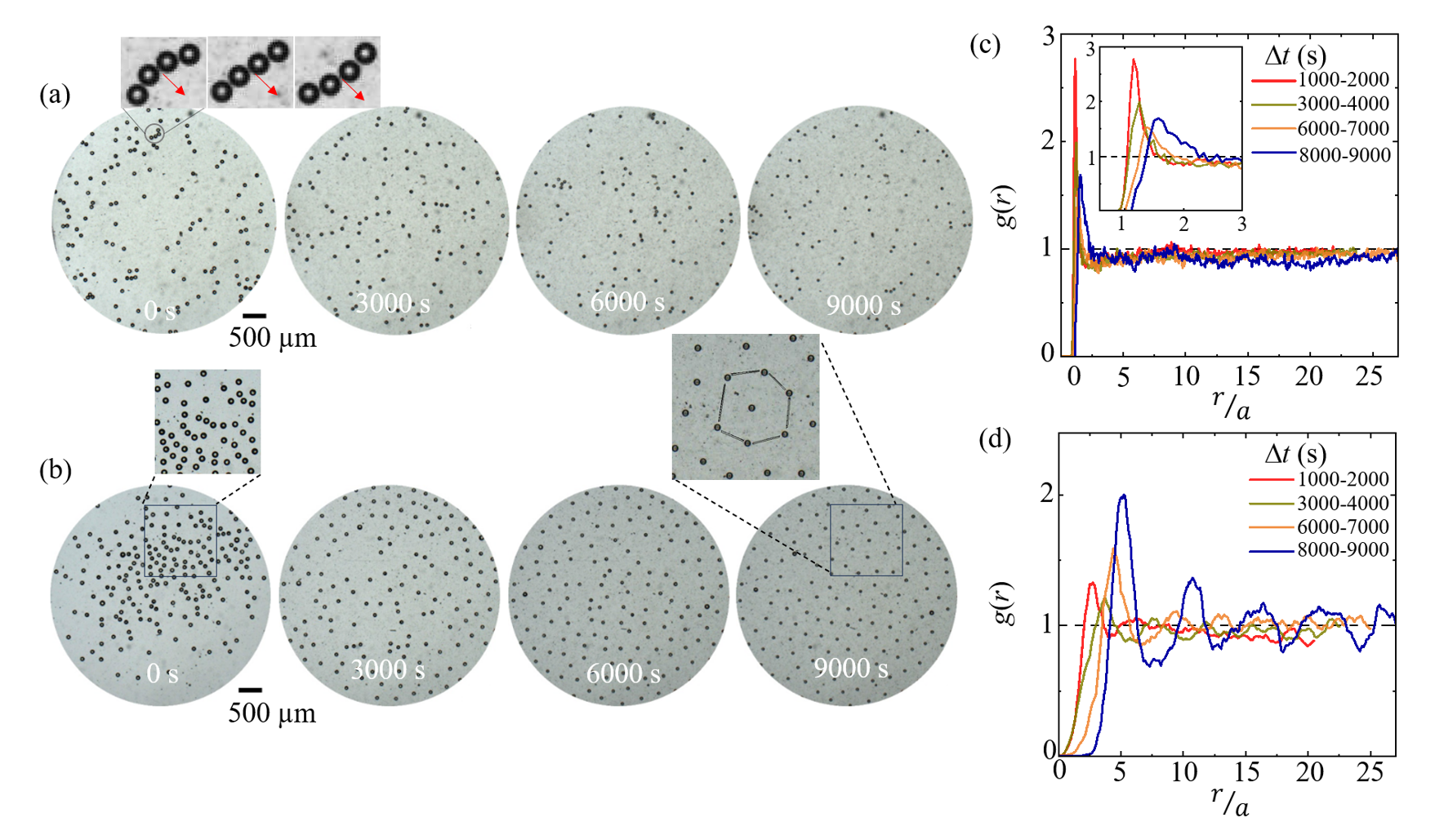}
    \caption{Optical micrographs captured at different time intervals depicting the collective motion of swimming 5CB droplets in a TTAB aqueous solution, with (a) no added solute and (b) PEO as an additive. The corresponding radial distribution functions (RDF) in (c) and (d) illustrate the spatial organization of droplets at distinct time intervals: 1000–2000 s, 3000–4000 s, 6000–7000 s, and 8000–9000 s. }
    \label{fig2}
\end{figure*}
%%%%%%%%%%%%%%%%%%%%%%%%%%%%%%%%%%%%%%%%%%%%%%%%%%%%

\subsection{Collective Motion of Droplets}

We first examined the collective behavior of $\sim$ 200 5CB active droplets in only 6 wt.$\%$ TTAB aqueous solution. For these droplets the initial (at $t$ $\sim$ 0 s) $Pe$ $\sim$ 60. Over time, a few of the droplets form dynamic clusters ranging in size from $\sim$ 2 to 4 (see supporting movie S1), leading to the formation of multiple lateral chain-like structures. The formation of such clusters is consistent with the side-ways hydrodynamic attraction between the pusher droplets \cite{thutupalli2018flow}. Within the clusters with size 3 or more, droplets positioned at the center of the chain propel faster than those at the ends. This difference in propulsion leads to a transition from a concave to a convex chain shape, eventually causing the dispersion of the droplets (refer to the zoomed-in insets in Figure \ref{fig2}a). This behavior, observed under two-dimensional confinement, closely aligns with the findings of Thutupalli {\it{et al.}} \cite{thutupalli2018flow}, where comparable chain formation was reported in 5CB+TTAB systems under similar confinement conditions. Over longer durations, while the droplets remain randomly distributed, their tendency to form such dynamic clusters diminishes with time, see supporting Figure S2, indicating a reduced likelihood of droplet pairs or clusters forming, a behavior which we will elaborate in upcoming section. Next, the collective behavior of $\sim$ 200 5CB droplets in a 6 wt.$\%$ TTAB solution in presence of 1 wt.$\%$ 8000 kDa PEO bearing $Pe$ $\sim$ 4 was investigated (see supporting movie S2). As depicted in the optical micrograph shown in Figure \ref{fig2}b, soon after the injection (at $t$ = 0 s), the droplets remain densely packed near the central region. Subsequently, with time, they gradually propagate radially outwards more from the outer region and less from the inner core (see supporting Figure S3). With further progression, by $t$ = 9000 s,  the droplets begin to self-organize into an ordered arrangement, each surrounded by six other droplets, forming a cage-like structure (see inset at $t$ = 9000 s).%Moreover, the droplets appear to be in jammed state  with very little mobility . 

To quantify and compare the order and packing of droplets in the two cases, the radial distribution function (RDF) $g(r)=\frac{\rho(r)} {\rho_{m}}$ was analyzed. Here local number density $\rho(r)$= $\frac{n}{2\pi rdr}$, where $n$ represents the number of droplets in a shell of thickness \textit{dr} at distance $r$ from the reference droplet, and $\rho_{m}$ = $\frac{N}{A}$ denotes the mean density of droplets, with $N$ being the total number of droplets present in an area $A$. For droplets in TTAB aqueous solution (without PEO), Figure 2c illustrates the radial distribution function \(g(r)\), averaged over all droplets for time intervals $\Delta$$t$ = 1000–2000 s, 3000–4000 s, 6000–7000 s, and 8000–9000 s. The results indicate that \(g(r)\) exhibits only short-range correlations for all $\Delta$$t$. Specifically, the first-nearest-neighbor peak occurs at a distance approximately equal to the droplet diameter (1\textit{a}), signifying that droplets predominantly form pairs. For larger distances, \(g(r)\) approaches 1, signifying an isotropic distribution typical of an unordered state. As the system evolves, the radial distribution function \(g(r)\) exhibits a similar trend. However, the height of the first peak decreases and shifts beyond (1\textit{a}). This observation aligns with our findings in Figure S2, where a reduction in attraction—reflected in the decreasing formation of clusters—is evident, and will be discussed in the subsequent sections.  % At larger distances, \(g(r)\) gradually approaches a value close to 1, showing an isotropic distribution in unordered state of the system. As the system evolves, a similar trend of \(g(r)\) is observed but the height of the first peak decreases and shifted at a value greater than (\(1\sigma\)), indicating a reduction in the probability of formation of droplet pairs or clusters. This behavior is attributed to increasing repulsive interactions between the droplets a consequence of decreasing Péclet number which are discussed later in detail. 
In contrast, as shown in Figure \ref{fig2}d, the introduction of PEO results in the development of distinct periodic peaks in the radial distribution function \(g(r)\) over time, with the peak heights progressively increasing. This suggests an increasing degree of spatial ordering in the arrangement of the droplets. Additionally, during the final stage ($\Delta$$t$ = 8000-9000 s) the first-nearest-neighbor peak gradually shifts to 5.3\textit{a}, indicating that the assembly does not follow a close-packing configuration. Additionally, subsequent peaks at radial distances of 10.8\textit{a}, 16.4\textit{a}, and 21\textit{a} further highlight the development of a consistent long-range ordered structure.
%\textcolor{red}{wherein 5.3 times the droplet diameter ($\sigma$ $\sim$ 60 $\mu$m) is the bond length of the resulting ordered structure}.\RM{Dont understand the purpose of the last line}

% Hence, the radial positions of peaks are close to the peaks that appear for a regular hexagonal crystal i.e.  5.3$\sigma$, $5.3\sqrt{3}$$\sigma$, 10.6$\sigma$, $5.3\sqrt{7}$$\sigma$, 15.9$\sigma$, which signifies a nearly hexatic crystalline-like state of the structure.} \RM{I thought that this was not the case. Didnt we discuss that you will write based on the ratio of the peak positions?}

%\textcolor{red}{This suggests that the mere presence of peaks is not sufficient for the system to be in an ordered state rather, the consistent replication of peaks across distinct sites is crucial}.\RM{dont understand this line aso}\textit{\textcolor{green}{It means the presence of peaks(as one can see in the rdf of water at last stages )in rdf does not imply the system to be ordered rather the order in which these peaks repeat is what decides the crystallinity.In case of water we see that there is no restriction on the occurence of these peaks..we see first prominant peaks at 2$\sigma$,4$\sigma$, etc but there is no trend of the peaks following the first ones.thus despite there were peaks in rdf which could mislead one to say that it is ordered but there is no correlation between these peaks.On the otherhand in case of PEO, we get peaks in rdf which occur after regular intervals}}

%%%%%%%%%%%%%%%%%%%%%%%%%%%%%%%%%%%%%%%%%%%%%%%%%%%%%%%%%%
%%%%%%%%%%%%%%%%%%%%%%%%%%%%%%%%%%%%%%%%%%%%%%%%
\begin{figure*}{}
    \centering
    \includegraphics[width=01\linewidth]{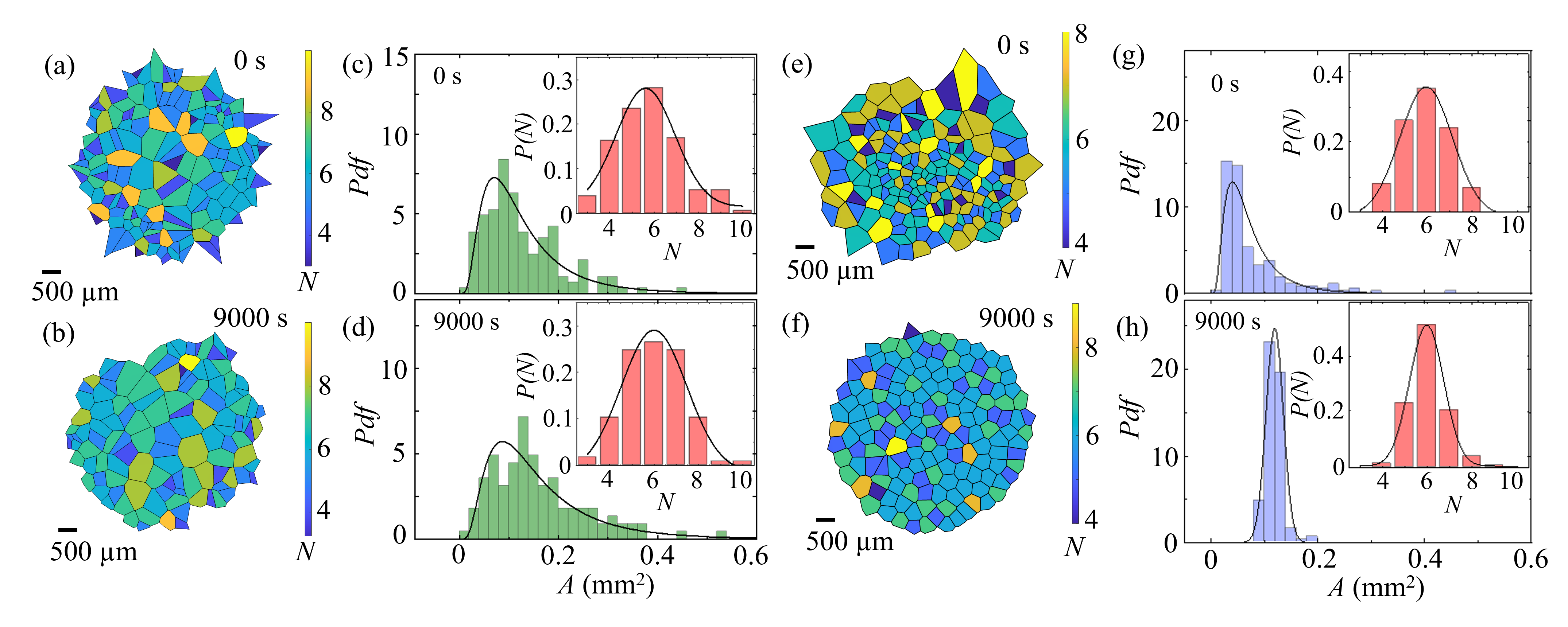}
    \caption{(a, b, e, f) Voronoi diagrams and the corresponding (c, d, g, h) probability density distributions of polygonal cell areas for 5CB droplets in a TTAB aqueous solution, examined under two conditions: (a–d) without any solute and (e–h) with PEO as an additive. The insets display histograms of the number of Voronoi cell edges along with their respective probabilities. The black curves represent the best-fit distributions.}
    \label{fig3}
\end{figure*}
%%%%%%%%%%%%%%%%%%%%%%%%%%%%%%%%%%%%%%%%%%%%%%%%%%%%%%
To gain deeper insights into the spatial distribution of droplets in both cases, we analyzed the Voronoi diagrams of the droplet patterns, as shown in Figures \ref{fig3}(a, b) and \ref{fig3}(e, f). Voronoi diagrams represent the partitioning of space into regions based on proximity to specific points, commonly referred to as ``seed" points \cite{Voronoi1908, lotito2020pattern}, in this case, the centroids of the droplets. This partitioning results in polygonal cells surrounding each seed point, with each cell encompassing the area closest to its corresponding seed point compared to any other point in the space. These cells provide valuable information on the nearest neighbor connections for each droplet and the areas they occupy \cite{lazar2022voronoi}.
In just the TTAB aqueous solution, the probability density distribution of the areas occupied by Voronoi cells (at $t$ = 0 and 9000 seconds), as depicted in Figures \ref{fig3}c and \ref{fig3}d, along with the corresponding histograms of the probabilities of polygonal cell edges (insets), show a broader range of distribution, indicating the unordered spread of droplets. However, with the addition of PEO, the initially disordered distribution of Voronoi cells becomes more uniform with time, transitioning from a broader distribution (Figure \ref{fig3}g) to a narrower one (Figure \ref{fig3}h). This shift signals the emergence of a long-range spatially ordered structure. The distribution of polygon edges becomes more concentrated around six edges (inset in Figure \ref{fig3}h), indicating that most droplets self-organize into a hexatic-like phase. However, due to fluctuations in droplet activity and positioning, not all droplets have exactly six neighbors, leading to deviations from a perfect hexatic arrangement. In fact, these deviations are expected to suppress the peaks at non integer multiples of 5.3$a$ (i.e. 5.3$\sqrt3$$a$, 5.3$\sqrt7$$a$) in the \(g(r)\) calculation, which is averaged over time and space. Consequently, only peaks at integer multiples of 5.3$a$ (i.e. 2(5.3$a$), 3(5.3$a$)) remain prominent, as observed in Figure \ref{fig2}d. Despite this, the system maintains long-range spatial order, underscoring the robustness of the emergent structure.

%peaks occurring at non-integer locations in the hexagonal arrangement of droplets when averaged over time and space. As a result, peaks corresponding to integer locations remain prominent. Nevertheless, the system exhibits long-range spatial ordered structure.}%From the area distribution at the time  t=9000s , approximating the mean area of voronoi cells (0.11 ${mm}^2$) to the area of a circle, we get the effective diameter ($d_{\text{eff}}$ = \sqrt{\frac{4a_{\text{mean}}}{\pi}} ) $\sim $380$\mu m$ which is close to the nearest neighbour distance ($320$ $\mu m$) calculated from the RDF thus assuring the system is nearly hexagonal ordered structure where in the bond length can be considered to be around 320-380 $\mu m$.
%%%%%%%%%%%%%%%%%%%%%%%%%%%%%%%%%%%%%%%%%%%%%%%%%%
\begin{figure*}
    \centering
    \includegraphics[width=1\linewidth]{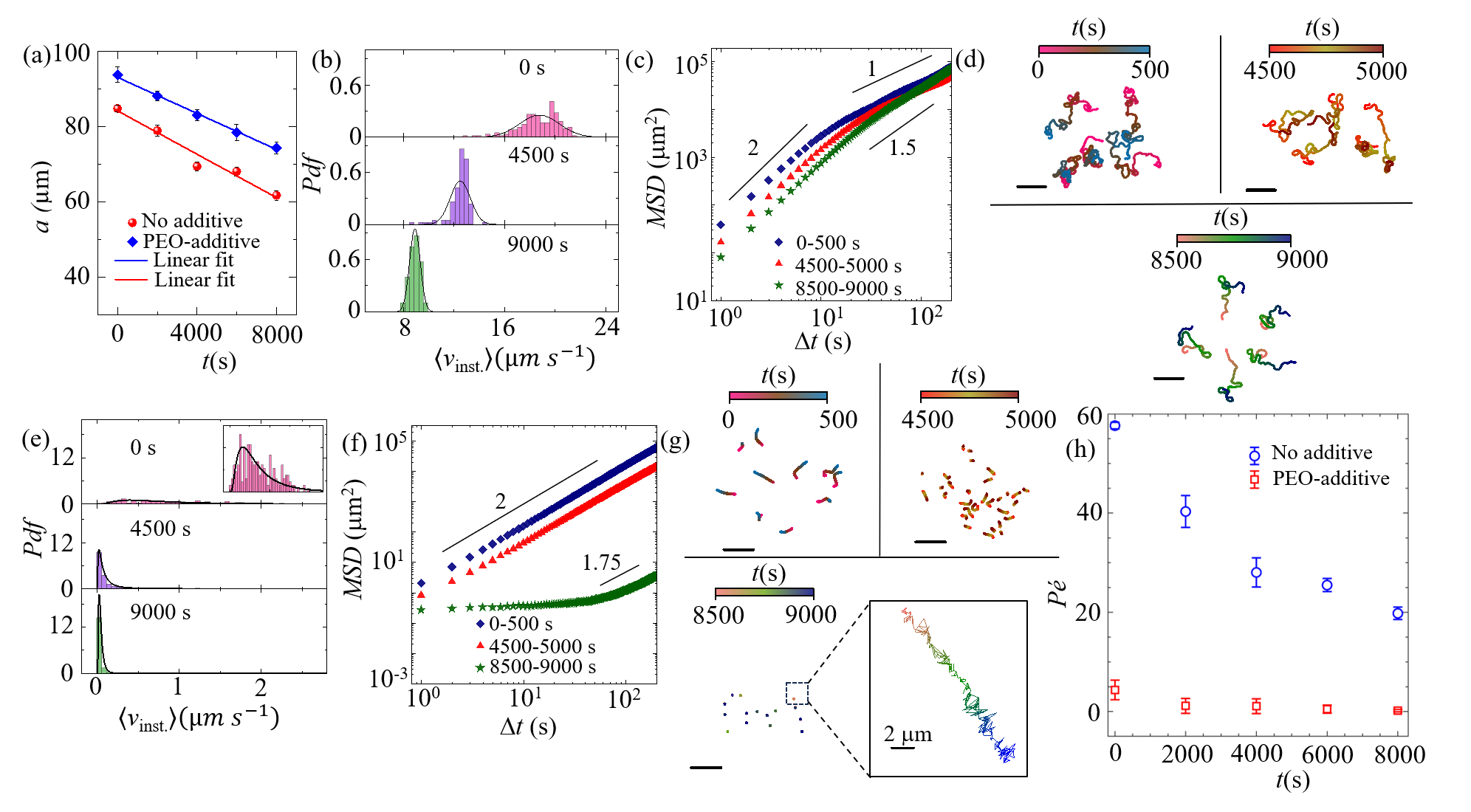}
    \caption{(a) Temporal evolution of droplet diameter in the absence and presence of PEO. For 5CB droplets swimming in pure TTAB aqueous solution (b),(c) and (d) are the mean instantaneous speed distributions, mean square displacement (MSD), and time evolution of droplets' $x$-$y$ trajectories, respectively. For 5CB droplets swimming in TTAB aqueous solution with PEO (e), (f) and (g) are the mean instantaneous speed distributions, mean square displacement (MSD), and time evolution of droplets' $x$-$y$ trajectories, respectively. (h) Temporal evolution of the Péclet number for both conditions. Scale bars in (d) and (g) represent 1000 $\mu$m.}
    \label{fig4}
\end{figure*}
%%%%%%%%%%%%%%%%%%%%%%%%%%%%%%%%%%%%%%
Further, in order to investigate the time-dependent dynamics of the droplets, we started with an assessment of the rate of solubilization of the droplets which is expected to significantly impact their swimming. As shown in Figure \ref{fig4}a, the average droplets' diameter decreases linearly in both the cases—without and with the addition of PEO to the TTAB aqueous solution—with slopes of -0.003 $\mu$m s$^{-1}$ and -0.0025 $\mu$m s$^{-1}$, respectively. This linear decrease in size aligns with previous studies \cite{peddireddy2012solubilization}, which attribute the constant rate of droplet shrinkage to micellar solubilization within a localized solubilization zone surrounding the droplets \cite{todorov2002kinetics}. The similar values of solubilization rates confirm that PEO does not affect the solubilization process. 

Figure \ref{fig4}b presents the probability distribution of the average speed of droplets (calculated as the mean instantaneous speed for each droplet over a 200 s interval) at three different time instants while swimming in absence of PEO. The distributions follow a normal fit, with a narrowing bandwidth and a decrease in mean speed from 19 $\mu$m s$^{-1}$ to 9 $\mu$m s$^{-1}$ with time. Figure \ref{fig4}c further characterizes the mean squared displacement (MSD) of the droplets over various time intervals. Each plot represents an average of the MSD from five distinct trajectories of swimming droplets over duration of 500 s. Across different times ($t$ = 0-500 s, 4500-5000 s, and 8500-9000 s), the droplets exhibit ballistic motion at short timescales, scaling as MSD $\sim$ $\Delta$$t$$^2$. However, at longer timescales, the MSD scaling undergoes significant changes as the system evolves. For  $t$ = 0-500 s, the long-time MSD scales as $\Delta{t}^{1}$ consistent with random swimming trajectories. As the system progresses to later stages $t$ = 8500-9000 s, the long-time scaling transitions to $\Delta{t}^{1.5}$. This evolution in scaling arises from the droplets' size reduction, which leads to reduced randomness in their motion (see Figure 4d).

In the presence of PEO as an additive, as discussed in the section B, the droplets in the central region remain crowded as opposed to those at the peripheral ones. This introduces a skewness in the velocity distributions evidenced as a lognormal speed distribution (see Figure \ref{fig4}e). Over time, the mode of the speed distribution shifts from 0.9 $\mu$m s$^{-1}$ at $t$ = 0 s to 0.04 $\mu$m s$^{-1}$ at $t$ = 9000 s, with a notable narrowing of the distribution width. The corresponding mean squared displacements (MSDs) reveal distinct dynamical behaviors across different time intervals. For the initial phase ($t$ = 0-500 s) and the mid-phase ($t$ = 4500-5000 s) MSD, as shown in Figure \ref{fig4}f demonstrates a consistent slope of 2 with respect to $\Delta{t}$, indicating persistent droplet motion (see Figure \ref{fig4}g) at all times. At the later stage ($t$ = 8500-9000 s), initially the MSD plateaus for an extended period, reflecting slower dynamics and caging behavior. At longer timescales, the MSD begins to increase with a slope of 1.75 (Figure \ref{fig4}f), indicating a superdiffusive behavior \cite{bi2016motility,Ganapathi2021structure}. However, despite the slope of 1.75, the MSD value remains quite low, reaching only around 10 $\mu$m$^2$ even for larger $\Delta$$t$. This implies that although the droplets exhibit superdiffusion behavior, the overall displacement is still constrained within a relatively small region due to their lower speeds and caging effect.

%This anomalous \textcolor{red}{subdiffusive} behavior is typical of motion in crowded/jammed environments consistent with  and extremely small trajectories\cite{bi2016motility,Ganapathi2021structure} The eventual increase in MSD at long time-scales distinguishes this ordering from that of solid crystals and suggests a resemblance to Wigner crystals, where long-range Coulomb interactions prevail over quantum fluctuations in electron movement \cite{li2021imaging}. Like Wigner crystals, the droplets remain separated, with the apparent confinement or cages being much larger (approximately 350 $\mu$m) compared to the droplet diameter.

Next, we aimed to assess the time-dependent variation of the Péclet number ($Pe$) in both scenarios. While the time evolution of droplet diameter (see Figure \ref{fig4}a) and corresponding speeds (see Figure \ref{fig4}b, e) is straightforward to measure, tracking the time evolution of micelle diffusivity ($D$) in such a collection of droplets is more complex. One potential cause for the change in $D$ could be the increase in bulk viscosity due to the rising concentration of filled micelles over time. To investigate this, we prepared a solution by dissolving 5CB oil in 6 wt.$\%$ TTAB in a 2:100 volume ratio, resulting in a filled micelle concentration similar to that expected in the droplet collection at the end of 9000s in either case. Upon comparison, we found no significant difference in the viscosities between the resultant solution ($\eta$ $\sim$ 1.6 mPa s) and the bare 6 wt.$\%$ TTAB solution ($\eta$ $\sim$ 1.4 mPa s). This observation suggests that the buildup of filled micelles during the experiment does not significantly alter the viscosity, and therefore, diffusivity ($D$) can be assumed to remain constant. Additionally, the time-invariant solubilization rate shown in Figure \ref{fig4}a supports the conclusion that $D$ remains time-independent.

To further ensure the accuracy of our estimates for diffusivity ($D$), we prepared another sample in which 5CB oil was dissolved in a 6 wt.$\%$ TTAB solution at a 1:100 ratio and then diluted to a 3:2 ratio with fresh 6 wt.$\%$ TTAB aqueous solution, yielding an overall concentration of 0.6 $\%$ v/v. This concentration of filled micelles represents the highest level at which self-propulsion of droplets could still be observed. The diffusivity of the released plumes in this micelle-rich solution was measured at approximately 25.8 $\mu$m$^2$ s$^{-1}$ (see supporting Figure S4(a, b)). This value is consistent with the diffusivity ($D$) observed in the pure 6 wt.$\%$ TTAB solution, further verifying that the bulk filled micelle concentration does not significantly influence the diffusivity and it can be considered constant throughout the process. Figure \ref{fig4}h illustrates the time evolution of the Péclet number ($Pe$) in both scenarios, calculated under this assumption. In the pure TTAB solution, the $Pe$ number decreases from 58 to 20 over 8000 seconds, while in the TTAB solution containing PEO, the $Pe$ number remains significantly lower and approaches nearly zero (from 4.3 to 0.18) within the same time interval.

These results highlight that the addition of PEO which significantly reduced the underlying $Pe$ of the swimming droplets, results in the emergence of dynamic out-of-equilibrium crystallization, where the droplets appear to be confined within a densely crowded environment. It also indicates that this behavior is mechanistically different from seemingly similar dynamic self-assembly observed in ensembles of camphor boats, which is attributed to the hydrodynamic interactions \cite{soh2008dynamic}.

%\textcolor{red}{Furthermore, PIV characterization around the swimming droplet and its tangential velocity fits along the droplet periphery shows the swimming mode $\beta>0$ (Figure S4(c, d)), a signature of puller mode.} \RM{This beta if for what system and What is the significance of the statement here?}\textcolor{red}{Sir this i don't think is much relevant now and i think we should not include it to inform that the droplet mode changes with increasing saturation} 
%a concentration equivalent to the volume of 200  droplets of 5CB of  initial Pe $\sim$$58$ (diameter $55$$\mu$ m) dissolving to $\sim$$20$ or(diameter $55$ $\mu$ m) in the given 9000s. 

%Since we observed a constant rate of solubilization independent of droplet size ,our system thus follows a rate limited micellar solubilization process in a narrow diffusion layer surrounding the droplet \cite{todorov2002kinetics} . 

\begin{figure*}
    \centering
    \includegraphics[width=1\linewidth]{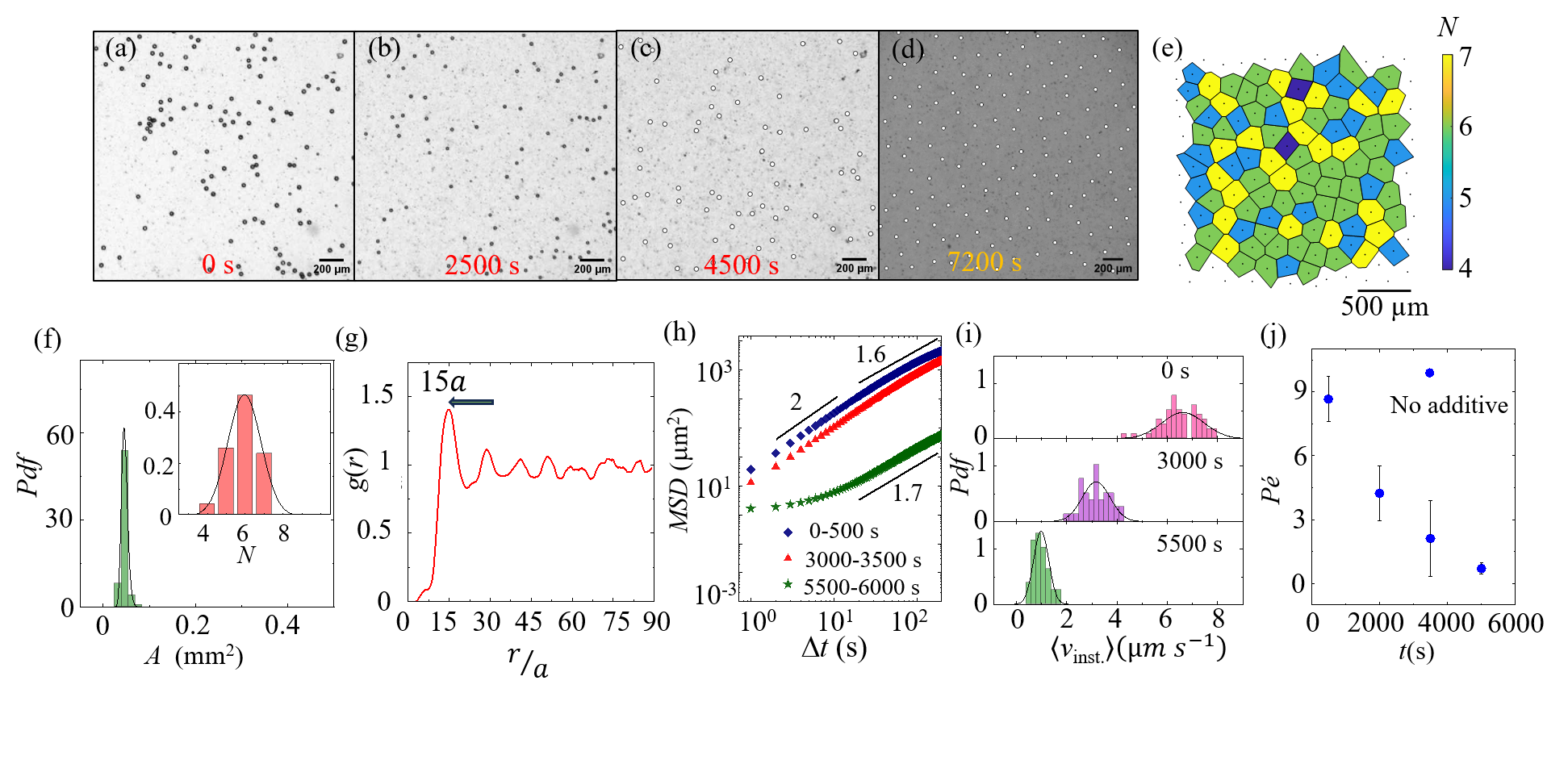}
    \caption{(a-d) Time-lapse optical micrographs of the droplets swimming in pure TTAB aqueous solution, beginning with an initial droplet diameter of 35 $\mu$m. In panels (c) and (d), the white dots do not indicate the actual size of the droplets but instead show their locations only. Panel (e) displays the Voronoi diagram at the final stage, with cells color-coded according to the number of sides (\textit{N}). (f) Probability density function of the area occupied by the Voronoi cells at the final stage. (g) Radial distribution function (RDF) of the system at the final stage, where an apparent ordering similar to the case of PEO is observed. (h) Mean squared displacements (MSDs) of droplet trajectories (each lasting 500 seconds) for three different designated time intervals. (i) Time evolution of instantaneous speeds (averaged over 200 seconds), demonstrating a normal distribution. (j) Evolution of the associated Péclet number ($Pe$) with time.}
    \label{fig5}
\end{figure*}

    \begin{figure}
    \centering
    \includegraphics[width=1\linewidth]{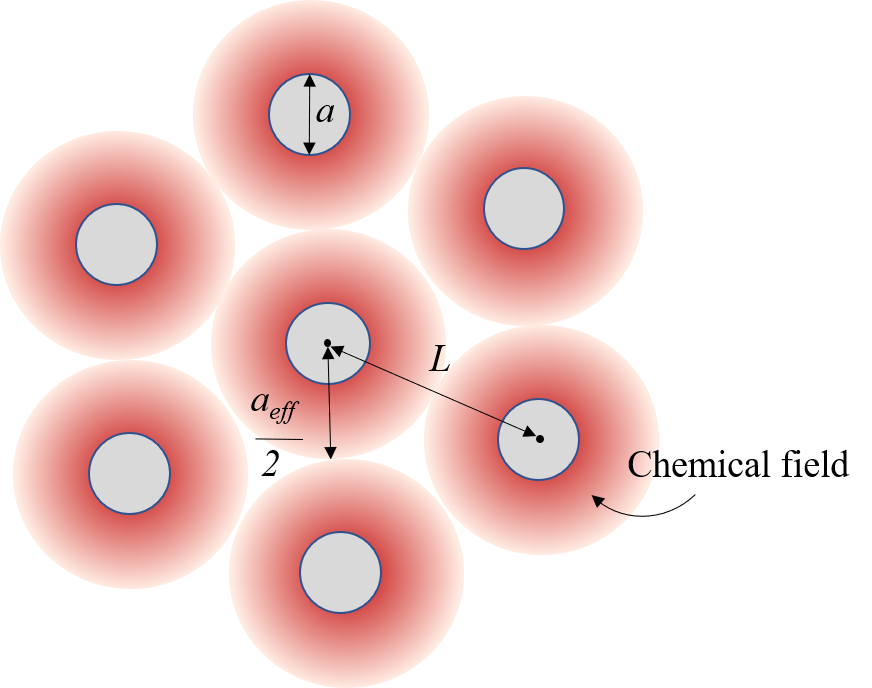}
    \caption{Schematic of the chemical field induced ordering in the active droplets.}

    \label{fig6}
\end{figure}

    %\subsection*{High  Péclet }
\begin{figure*}
    \centering
    \includegraphics[width=1\linewidth]{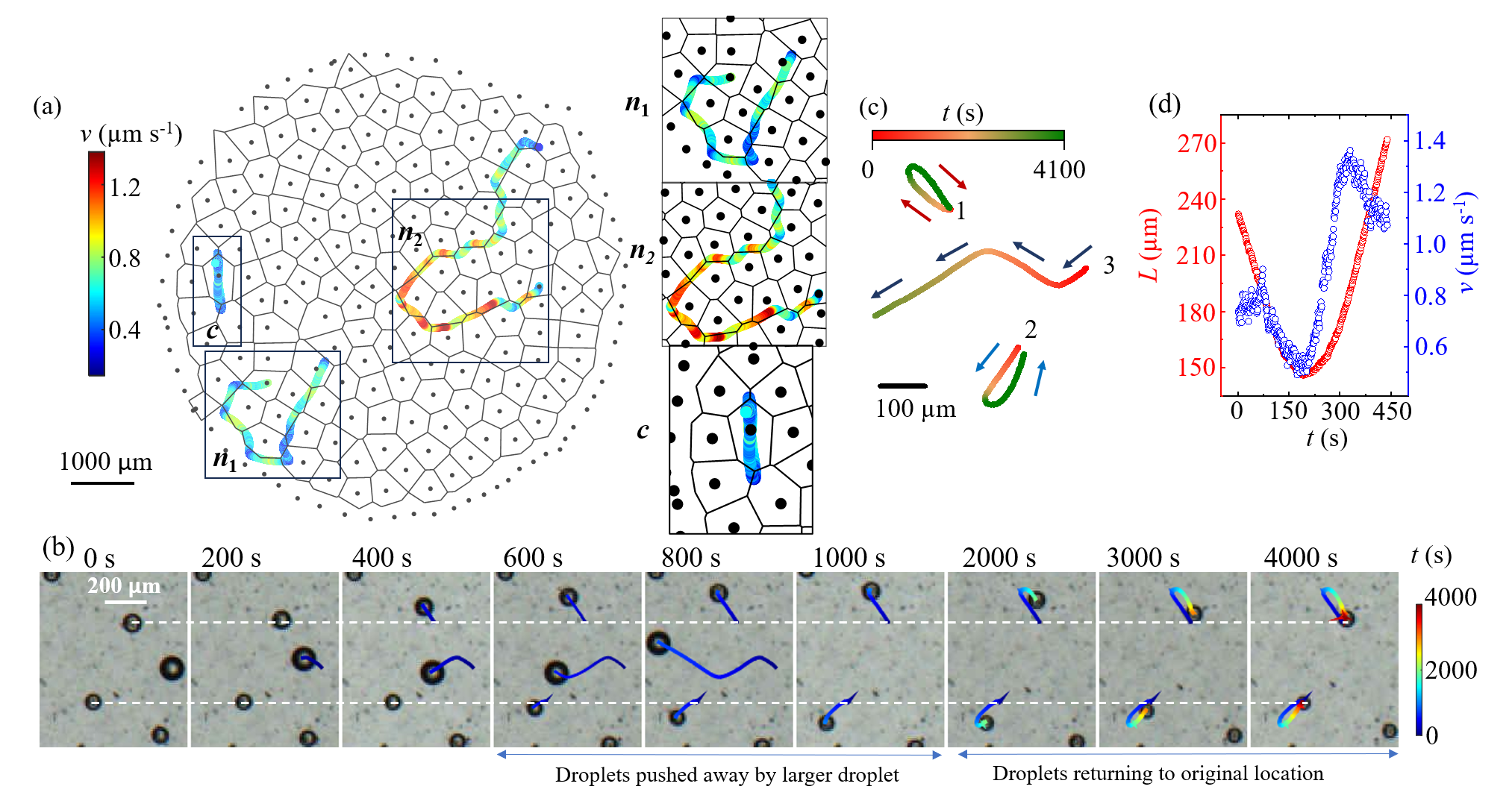}
    \caption{(a) Representative \(x-y\) trajectories of two navigating droplets (\(n_1\) and \(n_2\)) and a caged droplet (\(c\)), color-coded by speed and overlaid on the Voronoi diagram of a network of static droplets forming an ordered state. Panel (b) illustrates the time-lapse optical micrographs for the encounter of a navigating droplet with the caged droplets highlighting the softness of the cage. White dashed lines represent the reference positions of the caged droplets. (c) corresponding \(x-y\) trajectories of the droplets for the entire interaction. Here $\#$ 1 and $\#$ 2 represent the caged droplets, while $\#$ 3 represents the navigating larger droplet. Panel (d) depicts the variation in the inter-droplet distance ($L$) and the speed of a droplet, with time, as it interacts with a stationary droplet.} 
    \label{fig7}
\end{figure*}
%%%%%%%%%%%%%%%%%%%%%%%%%%%%%%%%%%%%%

\subsection{Crystallization in water }
To confirm that the presence of high molecular weight ($M_w$) polymer is not essential for droplet ordering and that it primarily depends on the droplets' underlying Péclet number ($Pe$), we aimed to achieve ordering in just the 6 wt.$\%$ TTAB aqueous solution. To significantly reduce $Pe$, 
we used smaller droplets, approximately 35 $\mu$m in size, and increased the droplet population to 600 to promote ordering. Initially, similar to experiments with larger droplets, clustering was observed. However, unlike larger droplets, this clustering gradually diminished over time, giving rise to dynamic crystallization, similar to that seen in the presence of PEO, as shown in Figures \ref{fig5}a–d. Micrograph \ref{fig5}d shows that after $\sim$ 7200 s droplets are spatially well-organized with an average separation of about 15\textit{a} (where \textit{a} represents the droplet diameter i.e. 15 $\mu$m at that stage) from their nearest neighbors (see supporting movie S3-S5). To analyze the spatial ordering, a Voronoi diagram was constructed (Figure \ref{fig5}e), and the probability density of the Voronoi cell areas (Figure \ref{fig5}f) revealed a narrow distribution, with majority of the cells being hexagonal (inset). The oscillating peaks of the radial distribution function, observed at regular intervals in Figure \ref{fig5}g, indicate a long-range ordered state, similar to what was observed in the case of PEO. For  $t$= 0-500 s and 3000-3500 s, the mean square displacement (MSD) for the droplets demonstrates a slope of 2 with respect to $\Delta{t}$ at short timescales which transitions to a slope of $\sim$ 1.6 at longer timescales (see Figure \ref{fig5}i). For $t$= 5500-6000 s, when the system reaches a dynamic ordered state, the MSD exhibits a plateau at short time scales, followed by an increase with a slope of $\sim$ 1.7. This behavior confirms that, similar to the case of PEO, in this ordered state, the droplets are dynamically trapped in an apparent crowded environment, where the cage size is larger than the droplet size (inter-droplet separation $\sim$ 15\textit{a}).

%\RM{I dont understand this line?} \textcolor{red}{Isolated swimming droplets of size $\sim$ 35 $\mu$m showed persistent motion \cite{dwivedi2021solute} however due to droplets' physio-chemical interactions superdiffusive scaling is observed over longer timescales. -----> We wrote this because, in the case of an isolated droplet scenario, the long-time scale MSD follows a $\Delta t^2$ scaling. However, for the multiple droplets system (the droplets suffer multiple reorientations owing to inter droplet collisions), we observed slope less than 2, which is expected due to chemical and flow-field interactions among the droplets.} 

The speed distribution illustrated in Figure \ref{fig5}i indicates that droplets experience a decrease in average speed from 6 $\mu$m $s^{-1}$ to 1 $\mu$m $s^{-1}$ over time. The reduction in droplet diameter (from 35 $\mu$m to 15 $\mu$m) combined with speed reduction resulted in a decrease in the Péclet number from approximately 9 to 0.7 over 6000 seconds (see Figure \ref{fig5}j). Notably, this lower $Pe$ range aligns with the values observed in the presence of PEO, confirming that dynamic ordering in active droplets can be achieved solely by reducing $Pe$ in simple aqueous systems, without the need for additional additives. These observations now provide a clear understanding of the decline in clustering over time for droplets swimming in an aqueous TTAB solution, as demonstrated in the earlier experiment (Figure \ref{fig2}a,c). With time, the decreasing underlying $Pe$ diminishes hydrodynamic effects while amplifying chemical interactions.

%\RM{Is it that due to lower size your Pe is even lower than the PEO case which is why the nearest neighbor is even far away?} \textcolor{red}{No, I think there might be some confusion due to the droplet size ($\sigma$). In the ordered state, the droplet size is 15 $\mu$m in water (distance is around 275 $\mu$m) and approximately 58 $\mu$m with the PEO additive (distance is around 360 $\mu$m). In both scenarios, we have used different population sizes so we can't compare.} 

\subsection{Reasons for $Pe$ induced dynamic self-assembly and $\phi$ dependence}
Chemically active droplets have been demonstrated to interact with its surroundings through the associated chemical fields \cite{moerman2017solute, meredith2020predator, lippera2020bouncing}. For micellar solublization based swimming droplets, significant variations in the underlying $Pe$ values are known to influence the distribution of the surrounding chemical field (field of filled micelles). At low $Pe$, the chemical field has been demonstrated to be more enveloping, whereas at higher $Pe$, it becomes predominantly concentrated near the rear side of the droplet \cite{dwivedi2024p}. This variation not only results in distinct swimming modes of the swimming droplet—characterized as puller/pusher modes, respectively—but can also significantly influence the interactions between the droplets. Recent experiments by Dwivedi \emph{et al.} \cite{P2024} demonstrated that at low $Pe$, droplets exhibit purely repulsive interactions regardless of their approach direction or orientation. This repulsion arises from the dominance of the chemical field. In contrast, at higher $Pe$, droplets display behaviors such as lateral coupling, chasing, and scattering, primarily governed by the hydrodynamic flow field surrounding them. These findings align with numerical studies by Liperra \emph{et al.}, which explored the chemical rebounding of active droplets in interactions with both walls and other droplets \cite{lippera2020collisions, lippera2021alignment}. These results suggest that the dynamic ordering behavior observed in our experiments at low $Pe$ (with or without PEO) can be attributed to repulsive chemical interactions and over time, driven by the underlying repulsive potential energy landscape, the system evolves into a dynamically ordered state, minimizing its free energy.

Decreasing $Pe$ allows chemical interactions to prevail over hydrodynamic ones, although droplet crystallization is also dependent on their number density, estimated by the area fraction ($\phi$). The supporting Figure S5 shows optical micrographs taken at $t$ = 9000 s for groups of swimming droplets with different $\phi$ in the presence of PEO. The comparison clearly indicates that while droplets remain separated at low $Pe$, ordering does not occur at low $\phi$ due to insufficient droplet density for effective interactions. Therefore, we argue that the emergence of an ordered state in the system requires both a critical $Pe$ and $\phi$ for which we propose the following criteria based on a simple back-of-the-envelope calculation. Long-range repulsive interactions can be considered to effectively enlarge the droplet to a size $a_{eff.}$. Ordering is expected to occur when the separation between droplets $L$, calculated as $a{\sqrt{\frac{\pi}{2\sqrt3\phi}}}$ under the assumption of a uniform hexagonal distribution, is less than $a_{eff.}$, as shown in Figure \ref{fig6}. This is in agreement with the fact that for the case of PEO where around 200 droplets ($a_{eff.}$ as effective diameter) are dispersed in a circular cavity of diameter 5.2 {$\pm$ 0.1} mm, the $L$ $\sim$ 340–360 $\mu$m, which closely aligns with the first nearest-neighbor peak (\(5.3 \textit{a} \sim 320\) $\mu$m) obtained from the radial distribution function (RDF). Likewise, for the case of ordering in just the aqueous TTAB solution, the calculated $L$ $\sim$ 200-205 $\mu$m, aligns well with the first nearest-neighbor peak (\(15\textit{a} \sim 225\) $\mu$m) from the RDF, further supporting the observed structural organization. Further, for interactions with walls, Liperra \etal demonstrated that at low $Pe$, the rebounding distance for active droplets scales as $\propto$ $Pe^{-1}$ \cite{lippera2020bouncing}. Therefore, considering $a_{eff.}$=$\frac{\psi}{Pe}$, where $\psi$ is a proportionality constant, the resulting criteria for ordering is if $Pe \phi^{-0.5} \le \psi$. Using numerical methods, a similar criterion for the stability of 2D phoretic disks was proposed by Yang \etal {\cite{yang2024shaping}. The study proposed that the boundary of the ordered-disordered region as $Pe_{col}$ = $C$, where $Pe_{col}$ is the collective $Pe$ number $\propto$ $Pe$$\phi^{-0.5}$. The constants ($\psi$ or $C$) are anticipated to be influenced by the specific nature of the droplet's interaction with the surrounding chemical field. Nevertheless, future experiments are warranted to confirm the proposed phase envelope. Overall, at low $Pe$, droplets interact via a long-range chemical field, leading to an ordered arrangement where confinement exceeds droplet size, akin to Wigner crystals dominated by long-range Coulomb interactions over electron quantum fluctuations \cite{li2021imaging}.

\subsection{Navigation in structured environment}
 
The self-organized pattern of droplets in an arrested state serves as a template to explore the navigation active droplets within a structured environment. To investigate this behavior, we deliberately introduced few larger 5CB droplets ($\sim$100 $\mu$m), serving as external swimmers, among the population of $\sim$ 200 droplets with a diameter of $d$ $\sim$ 80$\mu$m. Over a period of $\sim$ 9000 seconds, the smaller droplets reached a spatially ordered jammed state, as indicated by the droplets' centroids in Figure \ref{fig7}a. In contrast, the larger droplets remained active, navigating through the ordered structure within cages formed by the surrounding jammed droplets. The color-coded trajectories in Figure \ref{fig7}a (indicating speed) illustrate their motion. The Figure presents trajectories for three larger droplets, two of which—designated as $n_1$ and $n_2$—are navigating droplets for which the trajectories exhibit substantial overlap with the edges of the Voronoi cells. In the zoomed section of Figure \ref{fig7}a, illustrating droplets $n_1$ and $n_2$, the navigating droplets for the most of the times move along the midpoint between two neighboring droplets, aligning with the straight edges of the Voronoi cell. When their path is obstructed by a caged droplet, they deflect, with the deflection point precisely at the corner of the Voronoi cell. This suggests that, due to the repulsive chemical interactions, the droplets engage in a cautious, neighbor-avoiding walk (as shown in supporting movies S6 and S7). The third droplet, denoted as $c$, is confined within a ``cage"(see supporting movie S8). As it approaches the boundary, interactions between its chemical field and neighboring quiescent droplets induce a 180-degree reflection. A similar interaction occurs when the droplet reaches the opposite boundary, effectively preventing its escape. Consequently, the droplet remains trapped, exhibiting a periodic oscillatory motion. 
 
To better understand this behavior, we closely analyzed the dynamics of the navigating droplet as it traversed the cage. Time-lapse images of a section of the trajectory, color-coded to represent speed (shown in Figure \ref{fig7}b), highlight the softness of the cage, with the navigating droplet displacing the smaller droplets as it moves (see supporting movie S9). During this event, the navigating droplet also changes its direction. These displacements are driven by the repulsive interactions between the chemical fields of the smaller droplets and the navigating droplet, with the directions of displacement governed by the need to minimize the local free energy of the system. Once the larger droplet passes, the smaller droplets return to their original positions, as depicted by the representative $x-y$ trajectories in Figure \ref{fig7}c. Notably, for the navigating droplet, instances of lower speeds coincide with moments of direction change. This phenomenon occurs because, as demonstrated in Figure \ref{fig7}d, the separation distance, denoted as $L$, between the navigating droplet and the smaller droplets is minimal at these points. This proximity leads to maximum repulsive interactions, which substantially decreases the droplet's speed.

\section{Conclusion}

In this experimental investigation, using micellar solubilization-driven self-propelled 5CB droplets in an aqueous TTAB solution with high molecular weight ($M_w$) PEO as a solute, we demonstrate that at lower $Pe$ values, the chemical field surrounding the active droplets dominates, leading to stronger repulsive interactions. These interactions drive the droplets apart, resulting in a spatially organized structure with a nearly hexatic arrangement. In this long-range ordered state, all droplets remain caged while undergoing mild fluctuations around fixed positions. In contrast, at relatively higher $Pe$ values (in the absence of PEO), the droplets initially exhibit random motion and form dynamic chain-like clusters due to hydrodynamic interactions. These clusters propagate along straight paths before bifurcating after a few seconds. Over time, as the droplets shrink in size and the surrounding aqueous phase becomes enriched with filled micelles, their propulsion speed decreases, effectively reducing the $Pe$ value and ultimately transitioning the system into an ordered state. While the emergence of order requires a sufficiently low $Pe$, a critical number density ($\phi$) of droplets is also necessary, and a simple scaling analysis was proposed to establish this criterion.
We further demonstrated that the formation of a chemically structured environment, composed of anchored droplets, creates distinct permissible and restricted zones for externally introduced chemically active droplets. Despite the absence of cognitive abilities, the chemorepulsive interactions between the droplets effectively regulate their motion, guiding selective propagation through the interstitial spaces within the droplet network. By leveraging chemically mediated interactions, our findings provide a robust framework for dynamically tuning inter-particle interactions in active colloidal systems. Instead of relying solely on hydrodynamic coupling, designing and controlling localized chemical gradients can offer spatial and temporal control over clustering, phase transitions, and emergent self-organization. This could enable applications in programmable active materials and adaptive microrobotic swarms, where precise control over collective motion is required. From a materials perspective, such open, non-close-packed self-assembly can be particularly advantageous for optical and electrochemical sensing applications, where minimizing cross-talk between features is essential \cite{Isa2010particle}.\\

\section{acknowledgement}
We acknowledge the funding received by Department of Science and Technology (SR/FST/ETII055/2013), Science and Engineering Research Board (Grant numbers SB/S2/RJN105/2017 and CRG/2022/003763), Department of Science and Technology, India. 

\bibliography{ArXiv}

%{\LARGE \textbf{{Supporting Figures}}}
\renewcommand{\thefigure}{S\arabic{figure}} 
\setcounter{figure}{0} 

\begin{figure*}[]
{\LARGE \textbf{{Supporting Figures}}}
\centerline{\includegraphics[width=16cm]{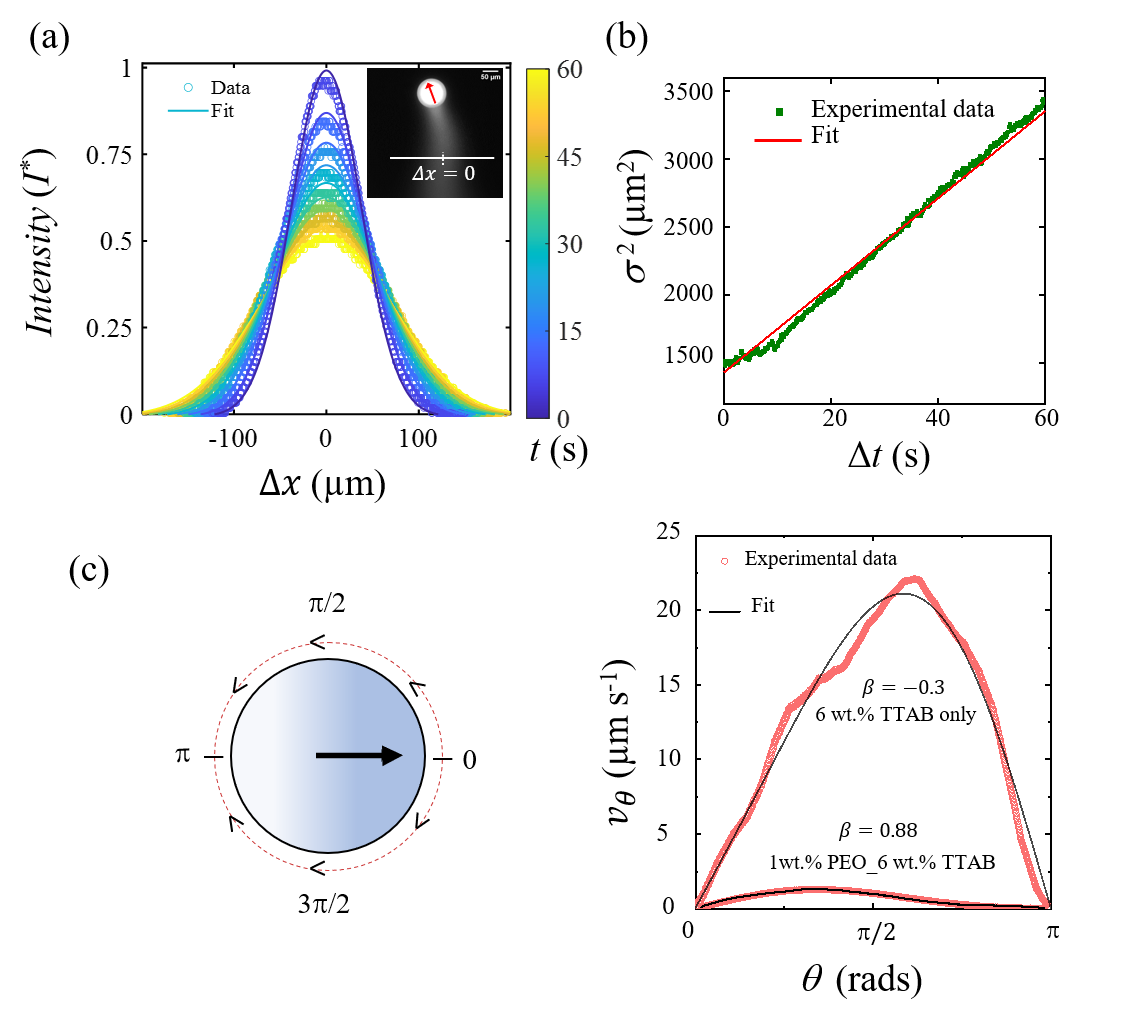}}
\linespread{1.0}
\caption{\small(a) The temporal evolution of the fluorescence intensity with overlaid guassian fits along the line taken perpendicular to the trail of the droplet (see inset showing the corresponding micrograph with red arrow pointing in the direction of propulsion) swimming in 6 wt.$\%$ aqueous TTAB solution. (b) The corresponding evolution of the standard deviation of the Gaussian fit with time where the solid line represents the linear fit. (c) Left panel is the schematic of the tangential velocity around a self-propelling droplet. Right panel illustrates both the experimental and theoretical fits for the asymptotic solution of the squirmer model. $u$($R$,$\theta$) = $B_1$$\sin(\theta)$+$\frac{B_2}{2}$$\sin(2\theta)$ results in squirmer parameter $\beta$ = $\frac{B_2}{B_1}$ =  -0.3 and 0.88, referring to a pusher and puller swimming mode in 6 wt.$\%$ aqueous TTAB solution without and with 1wt.$\%$ PEO (8000kDa), respectively.}
\label{S1}
\end{figure*}

\begin{figure*}[h]
\centerline{\includegraphics[width=7cm,height=7cm]{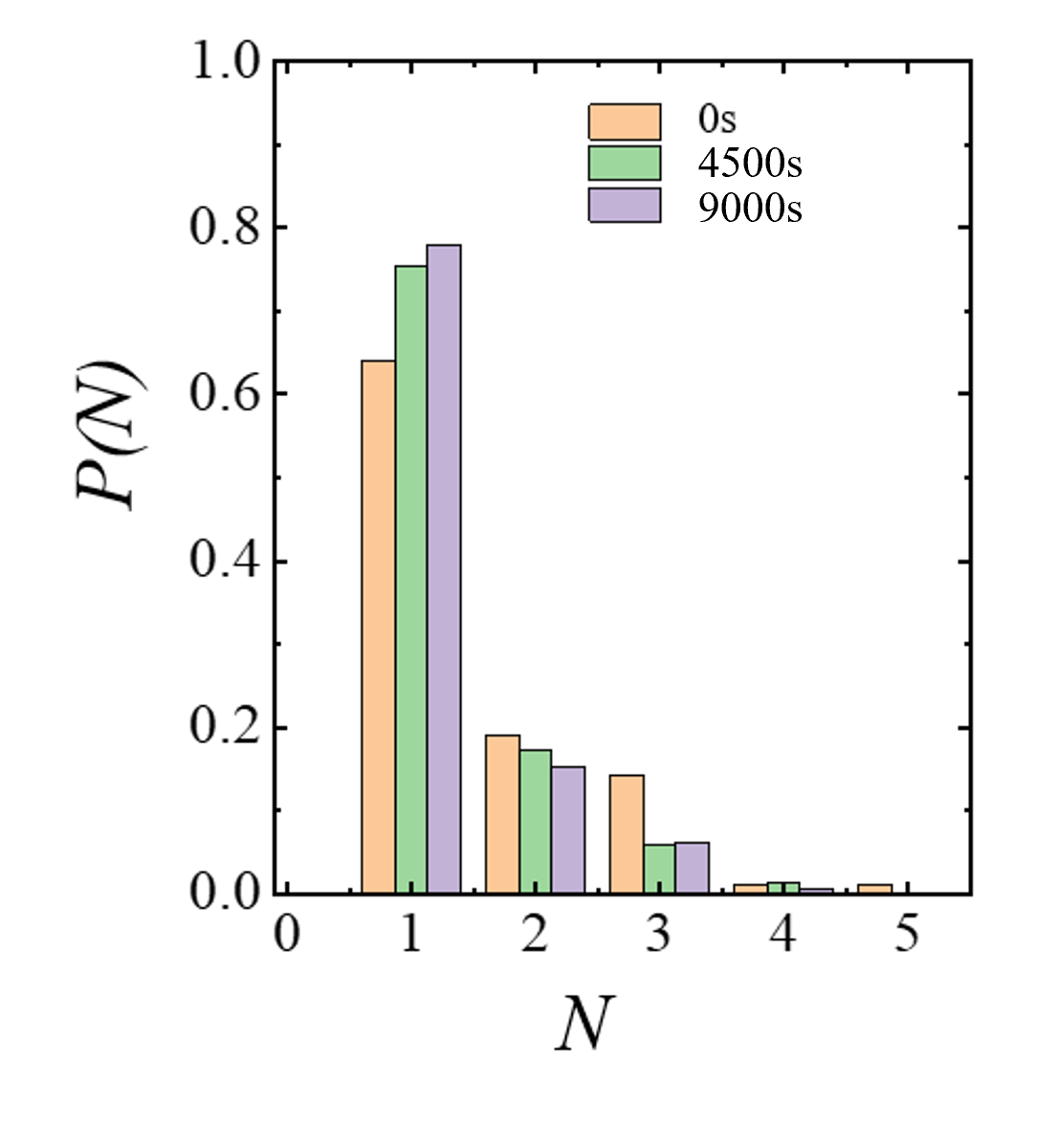}}
\linespread{1.0}
\caption{\small The temporal evolution of \textit{P(N)} the probability of encountering a cluster with \textit{N} droplets at $t$ = 0s, 4500s, and 8500s, demonstrating a decline in the frequency of clusters with $N$.}
\label{S2}
\end{figure*}

%%%%%%%%%%%%%%%%%%%%%%%%%
\begin{figure*}[h]
\centerline{\includegraphics[width=17cm]{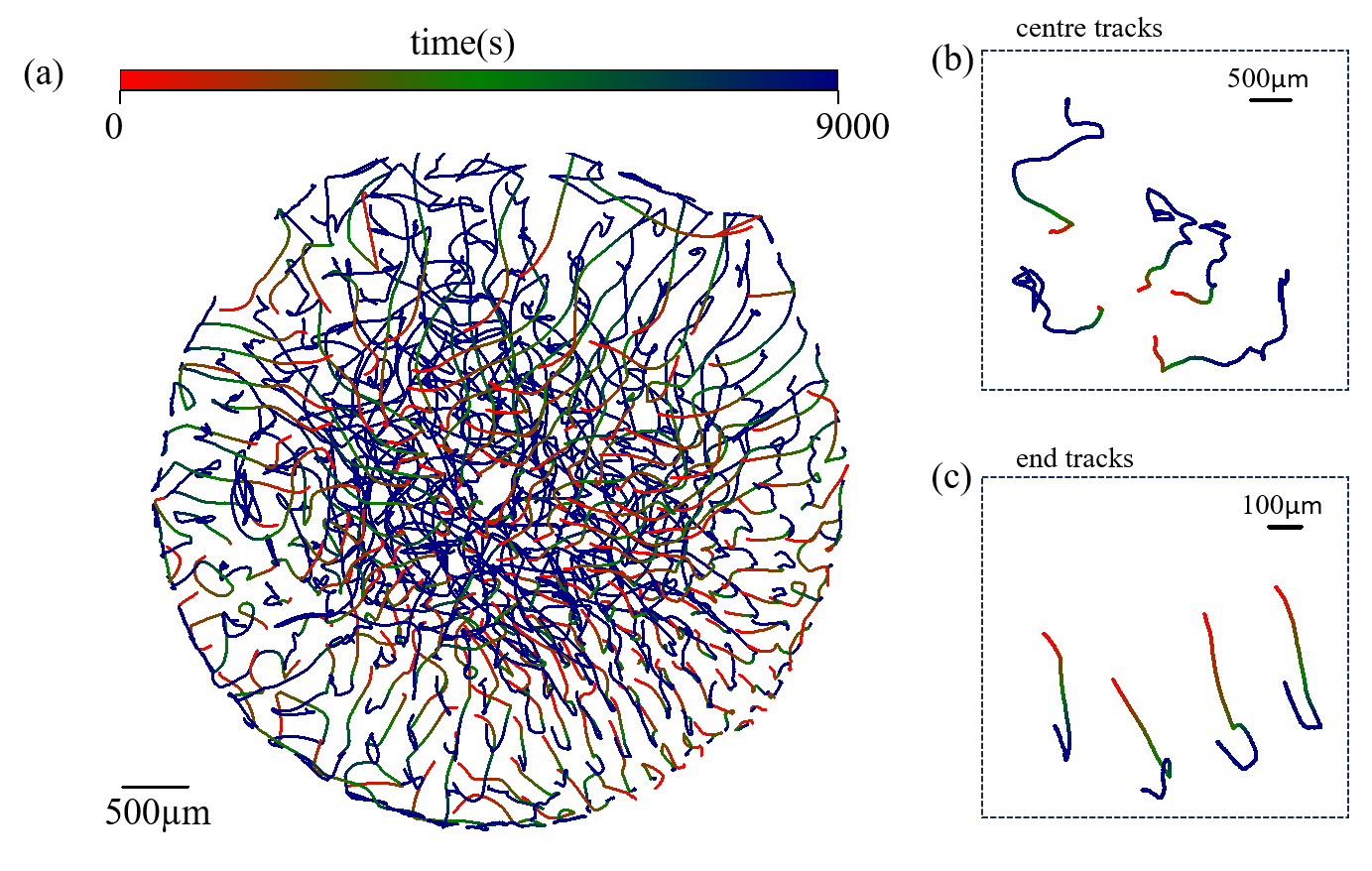}}
\linespread{1.0}
\caption{\small (a) Representative $x-y$ trajectories of the droplets in aqueous TTAB solution with PEO as additive. Panels (b) and (c) display the trajectories of droplets in the central and peripheral regions, respectively with droplets in the central region displaying multiple reorientations, while droplets at the edges moving radially outward accompanied with minimal reorientation events.} 
\label{S3}
\end{figure*}
%%%%%%%%%%%%%%%%%%%%%%%%
%%%%%%%%%%%%%%%%%%%%%%%%%%%%%%%%%%%%%%%%%%%%%%%%%%

%%%%%%%%%%%%%%%%%%%%%%%%%%%%%%%%%%%%%%%%%%%%%%%%%
%%%%%%%%%%%%%%%%%%%%%%%%%
\begin{figure*}[h]
\centerline{\includegraphics[width=17cm]{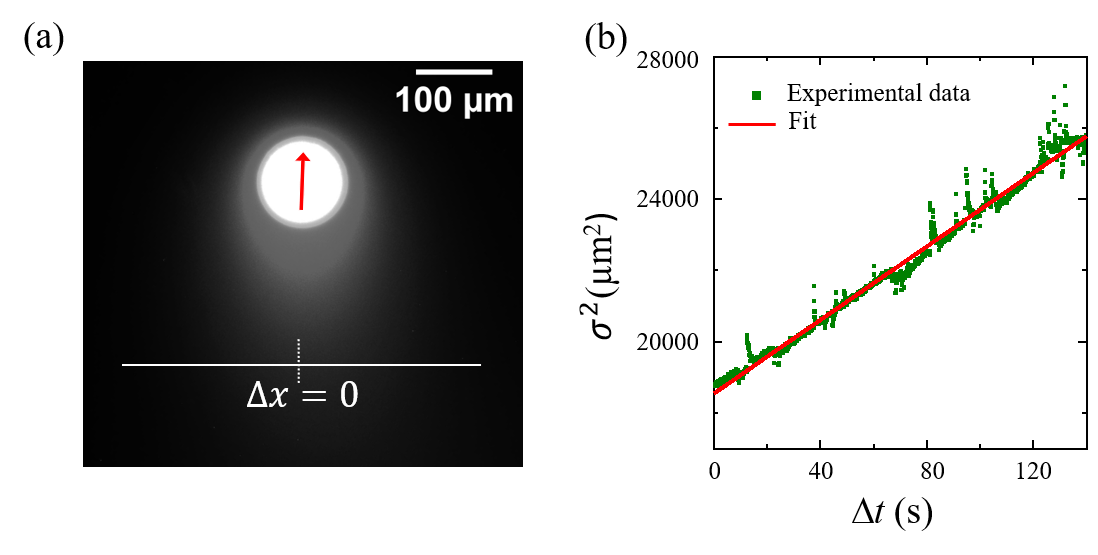}}
\linespread{1.0}
\caption{\small (a) Fluorescence micrograph demonstrating an active 5CB droplet moving in the direction marked by red arrow within a chemically adjusted environment, which includes an oil-filled micellar solution. (b) The temporal changes in the standard deviation of Gaussian curve fits to the fluorescence intensity along a line perpendicular to the droplet’s path, measured at a distance of 4$R$ from the droplet.}
 
\label{S4}
\end{figure*}
%%%%%%%%%%%%%%%%%%%%%%%%%
%%%%%%%%%%%%%%%%%%%%%%%%%%%%%%%
%%%%%%%%%%%%%%%%%%%%%%%%%
\begin{figure*}[h]
\centerline{\includegraphics[width=17cm]{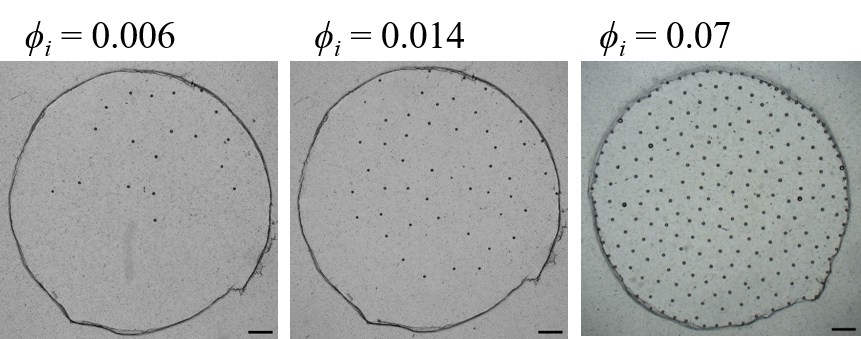}}
\linespread{1.0}
\caption{\small Bright field micrographs for the arrangement of droplets after 9000 seconds in the presence of PEO highlighting the role of initial area fraction ($\phi_{i}$) in promoting ordering at a given $Pe$. Scale bar = 500 $\mu$m.  }
\label{S5}
\end{figure*}
%%%%%%%%%%%%%%%%%%%%%%%%%
%%%%%%%%%%%%%%%%%%%%%%%%%%%%%%%
%%%%%%%%%%%%%%%%%%%%%%%%%
\clearpage
\begin{center}
    \LARGE \textbf{Supporting Movies}
\end{center}

\begin{description}
\item[Movie S1]TrackMate capture of Collective behaviour of 200 droplets of 5CB droplet in a quasi 2D cell in 6 wt$\%$ TTAB aqueous solution, shown at 100 X the actual speed.
\item[Movie S2]TrackMate capture of Collective behaviour of 200 droplets of 5CB droplet in a quasi 2D cell in 1 wt$\%$ PEO aqueous TTAB solution, shown at 100 X the actual speed.
\item[Movie S3]TrackMate capture of the collective behaviour of droplets of size 35$\mu m$ in  6 wt$\%$ TTAB aqueous solution for $t=500-1000s$. Movie is shown at 20 X the actual speed.
\item[Movie S4]TrackMate capture of the collective behaviour of droplets of size 35$\mu m$ in  6 wt$\%$ TTAB aqueous solution for $t=3000-3500s$. Movie is shown at 20 X the actual speed.
\item[Movie S5] TrackMate capture of the collective behaviour of droplets of size 35$\mu m$ in  6 wt$\%$ TTAB aqueous solution for $t=5500-6000s$. Movie is shown at 20 X the actual speed.
\item[Movie S6] Navigating droplet $n_1$ as it makes its way through the droplet assembly. Movie is shown at 800 X the actual speed.
\item[Movie S7] Navigating droplet $n_2$ as it makes its way through the droplet assembly. Movie is shown at 800 X the actual speed.
\item[Movie S8] A caged droplet $c$ rattling in a cage formed by surrounding neighbours. Movie is shown at 800 X the actual speed.
\item[Movie S9] TrackMate capture a droplet displacing the surrounding droplets, with the droplets eventually returning to their original positions thereby demonstrating the softness of the cage. Movie is shown at 400 X the actual speed.
\end{description}
%\end{acknowledgement}

%%%%%%%%%%%%%%%%%%%%%%%%%%%%%%%%%%%%%%%%%%%%%%%%%%%%%%%%%%%%%%%%%%%%%
%% The same is true for Supporting Information, which should use the
%% suppinfo environment.
%%%%%%%%%%%%%%%%%%%%%%%%%%%%%%%%%%%%%%%%%%%%%%%%%%%%%%%%%%%%%%%%%%%%%\end{suppinfo}

%%%%%%%%%%%%%%%%%%%%%%%%%%%%%%%%%%%%%%%%%%%%%%%%%%%%%%%%%%%%%%%%%%%%%
%% The appropriate \bibliography command should be placed here.
%% Notice that the class file automatically sets \bibliographystyle
%% and also names the section correctly.
%%%%%%%%%%%%%%%%%%%%%%%%%%%%%%%%%%%%%%%%%%%%%%%%%%%%%%%%%%%%%%%%%%%%%

%\begin{singlespace}

\end{document}